\documentclass{aa}
\usepackage{graphics}
\usepackage{epsfig}

\def\msol{M_\odot}

\def\simgr{\,\hbox{\hbox{$ > $}\kern -0.8em \lower 1.0ex\hbox{$\sim$}}\,}
\def\simle{\,\hbox{\hbox{$ < $}\kern -0.8em \lower 1.0ex\hbox{$\sim$}}\,}
\def\beq{\begin{equation}}
\def\eeq{\end{equation}}
\def\pmin{P_{\rm min}}
\def\pturn{P_{\rm turn}}
\def\araa{ARA\&A}             
\def\apj{ApJ}                 
\def\aap{A\&A}                
\def\mnras{MNRAS}             

\begin{document}


\title{Distortion of secondaries in semi-detached binaries and the
 cataclysmic variable period minimum}
 \author{V. Renvoiz\'e\inst{1},  I. Baraffe\inst{1,2}, U. Kolb\inst{3}
\and 
H. Ritter\inst{2}
}

\offprints{V. Renvoiz\'e}

\institute{C.R.A.L (UMR 5574 CNRS),
 Ecole Normale Sup\'erieure, 69364 Lyon
Cedex 07, France (vrenvoiz, ibaraffe@ens-lyon.fr)
\and Max-Planck Institut f\"ur Astrophysik, Karl-Schwarzschild-str.1,
D-85741 Garching, Germany (hsr@mpa-garching.mpg.de)
\and
Department of Physics \& Astronomy, The Open University, Walton Hall,
Milton Keynes, MK7~6AA, UK (u.c.kolb@open.ac.uk)
}

\date{Received /Accepted}

\titlerunning{Distortion of secondaries in semi-detached binaries}
\authorrunning{Renvoiz\'e et al.}
\abstract{Based on SPH simulations,
we quantify the geometrical distortion effect due to tidal and rotational
forces on polytropic secondaries in semi-detached binaries. 
The main effect is an expansion of the polytropic star, with an effect
on the radius of $\sim$ 5\%-12\%,
depending on the
polytropic index  and the mass ratio. We apply such distortion
effects to the secular evolution of secondaries  in
cataclysmic variable systems. We focus on systems below
the 2-3h period gap and that approach 
the minimum period. We find
a significant increase of the predicted minimum period ($\sim$
4\% if changes in the secondary's thermal relaxation are approximately
taken into account).
Though an improvement, the effect is not big enough to 
solve the mismatch between predicted and observed minimum period
at 80 min.
\keywords{ binaries: close --- stars: low-mass, 
brown dwarfs --- stars: evolution --- cataclysmic variables}
}

\maketitle

\section{Introduction}

The description of close binary systems is usually based on the Roche
 model which defines the shape of a binary component distorted by
tidal and rotational forces. In the framework of  the Roche
model  one assumes that the binary
components (the primary and the secondary) either are point
masses, or are corotating and have a spherically symmetric mass
distribution irrespective of their proximity or mass ratio (Kopal
1959, 1978). In a semi--detached system, one of the components fills
its critical equipotential lobe defined by the  potential of the
inner Lagrangian point, and which determines the maximum extent of a
star in a close binary. This is the so--called Roche lobe within the
Roche model. Cataclysmic variables (hereafter CVs), composed of a
white dwarf as the primary and a low--mass star or a brown dwarf as
the secondary, belong to this type of systems: the secondary fills its
critical lobe and transfers mass towards the primary.  When
applying the Roche model to problems of binary evolution one makes
implicitly the following assumptions (among others). First, the
Roche potential is a good approximation of the true potential that
one would obtain by solving the Poisson equation. Second, the
effects of tidal and rotational forces on the internal structure of
the star are negligible, i.e. that they result in only small
corrections compared to stellar models assuming spherical geometry.
Third, that for the purpose of evolutionary computations involving
one--dimensional stellar models the lobe--filling star may be
replaced by a spherical star of the same volume. This is
tantamount to assuming that though tidal and rotational forces change
the shape of a star they leave its volume invariant. The radius of the
lobe--filling star then only depends on the geometry of the
system, and can be calculated by means of simple
analytical fits (Paczy\'nski 1971, Eggleton 1983). The main
purpose of the present paper is to examine in some detail the third
and to some extent also the second of the above assumptions, both of
which have so far not been tested.

Recent 3D simulations (Rezzolla et al. 2001; Motl \& Frank, priv. comm.)
confirm that, at least in the case of a
semi-detached system, the Roche potential is a good approximation if the
lobe-filling star is sufficiently centrally condensed, i.e. if the
effective polytropic index is $N \simgr$ 3/2.
The analysis of Rezzolla et al. (2001)
is based on numerical models of semi-detached binaries
that account for the finite size of the secondary star,
thus relaxing the first assumption inherent in the Roche model.
With the validity of this approximation for the determination of
the potential, they also show 
that such effects hardly affect gravitational quadrupole
radiation. 
Moreover, a comparison between the angular momentum loss and mass-transfer
timescales predicted by the Roche model and their numerical models
shows small differences. 
They thus conclude that finite
size effects cannot account for the mismatch between the observed
minimum period $\pmin$ at 80 min of CV systems and the theoretical value
$\pturn$.
The latter is indeed $\sim$ 15\% shorter than the observed
value, according to recent
calculations based on improved stellar physics (see Kolb \& Baraffe
1999). Since Rezzolla et al. (2001) do not consider thermal relaxation
effects in their calculations, they  can only determine a
differential correction to
$\pturn$ when going from Roche model to self-consistent potential.
However, in doing so, the second  and third 
assumptions mentioned above
remain untested. The main purpose of our paper
is thus to explore the consequences of making these two assumptions.

Our main goal is to determine quantitatively
the departure from spherical symmetry of the secondary in
semi-detached
binaries
and to analyse the consequences on the mass transfer rates 
and orbital period in CV systems. We use smoothed particle
hydrodynamics (SPH)
techniques to study equilibrium configurations 
of semi-detached
binaries and estimate for different mass ratios the
geometrical deformation of the secondary as it fills its critical lobe.
The numerical models and results
are described in \S 2. In \S 3, we analyse some of the
consequences 
 of the tidal and rotational forces on
the secular evolution of the low mass donor on grounds of
 models constructed by  Kolb \& Baraffe (1999) and
Baraffe \& Kolb (2000).
We focus on the  problem of the minimum period and the
discrepancy between observations and models (see e.g
King 1988 and Kolb 2001 for a review on the properties of CV systems).  
A discussion and conclusions follow in \S 4.  

\section{Numerical models of semi-detached binary systems}

\subsection{The method}

We  used a SPH code,
originally developed and kindly provided 
by Willy Benz (see details in Benz et al. 1990)
to perform numerical simulations of a close binary
system composed of a point mass (primary) and a polytropic
star (secondary). The SPH method  has been  described extensively in the
literature (see  Monaghan 1992 and references therein) 
and  is often applied to the study of close binary systems (e.g Benz
et al. 1990;
Lai et al. 1994; Rasio \& Shapiro 1995; Segr\'etain et al. 1997)

In all our simulations, the primary is a 1 $\msol$ point-like mass.
The secondary is described by a polytropic equation of state
 $p = K \rho ^{1+1/N}$,  where $p$ is the pressure and $\rho$ the
density. The polytropic constant $K$ is fixed for a given index
$N$ by the mass $M$ and radius $R$ of the  spherical secondary.  
We adopt two polytropic indices, i.e. $N$ = 3/2, which provides a good
description of fully convective objects such as low mass stars, and $N$ = 3
characteristic of solar type stars with $M \sim 1 \msol$.
For the particular case of CVs, systems below the period gap
are well described by $N$ = 3/2 polytropes, whereas $N$ = 3
applies to systems with periods $>$  6 h and typical masses 
around $\sim$ 1 $\msol$. The two values of $N$ thus represent
limiting cases for the description of CV secondaries.
Assuming that $K$ remains constant in space applies well
to fully convective objects with a fully adiabatic structure,
and implies a chemically homogeneous structure (constant molecular
weight) for the standard models with $N$ = 3. This is a reasonable
approximation for the present study.

The  simulations use $\sim$ 15000 particles. 
In order to check the accuracy of our results, we ran a limited number
of simulations with 57000 particles. We find that
15000 particles is a good compromise between computational 
demand and accuracy.
The particles are initially
uniformly distributed on a  hexagonal close-packed lattice. 
The initial number density of particles is constant
throughout the volume of the sphere describing the
initial configuration of the secondary. The particle masses
are proportional to the local mass density. This
provides a good spatial resolution near the stellar surface,
which is crucial for our problem of critical 
lobe determination where surface effects are predominant.
The simulations are performed in a corotating reference frame
with the origin at the center of mass of the system. The initial
separation $A_{\rm init}$ 
of the two components is arbitrarily fixed at four times
the separation required for the secondary to fill its
Roche-lobe $A_{\rm Roche}$, estimated from the Eggleton (1983) fit. For such a
separation, tidal and rotational 
effects on the secondary  are negligible.
The orbital separation is decreased with the arbitrary constant rate
$(A_{\rm init} - A_{\rm Roche})/\tau_{\rm simu}$,
so that the total timescale of the simulation $\tau_{\rm simu}$ is $\sim$
1000 times the typical hydrodynamical relaxation time 
 $\tau_{\rm relax} \simeq \left(\frac{R^3}{GM}\right)^{1/2}$
of the secondary.
The simulation is stopped when the secondary fills its
critical lobe {\it i.e.} when the first particles from the secondary
reach the inner saddle point of the potential \footnote{defined as $L_1$
in the Roche potential}.
This marks the onset of mass transfer.
Once the critical separation is reached, we check that the model
has reached an equilibrium configuration, starting
from such critical separation and letting it relax
in a non rotating reference frame.

Our goal is to estimate the deformation effects on the secondary due
to tidal and rotational forces as it fills its critical lobe.  The
deformation can be measured in terms of the ratio of the final to
initial stellar radius $D = R_{\rm f}/R_{\rm i}$.
$R_{\rm i}$ is the radius of the unperturbed spherical polytrope.
$R_{\rm f}$ is an effective radius defined as the radius of the sphere
with the volume $V_{\rm f}$ of the secondary filling its critical lobe. 
$V_{\rm f}$ is
provided by our SPH simulation at the onset of mass transfer. 
The method to estimate $V_{\rm f}$ is described in Appendix A.
  
\subsection{Results}

\begin{figure*}
\psfig{file=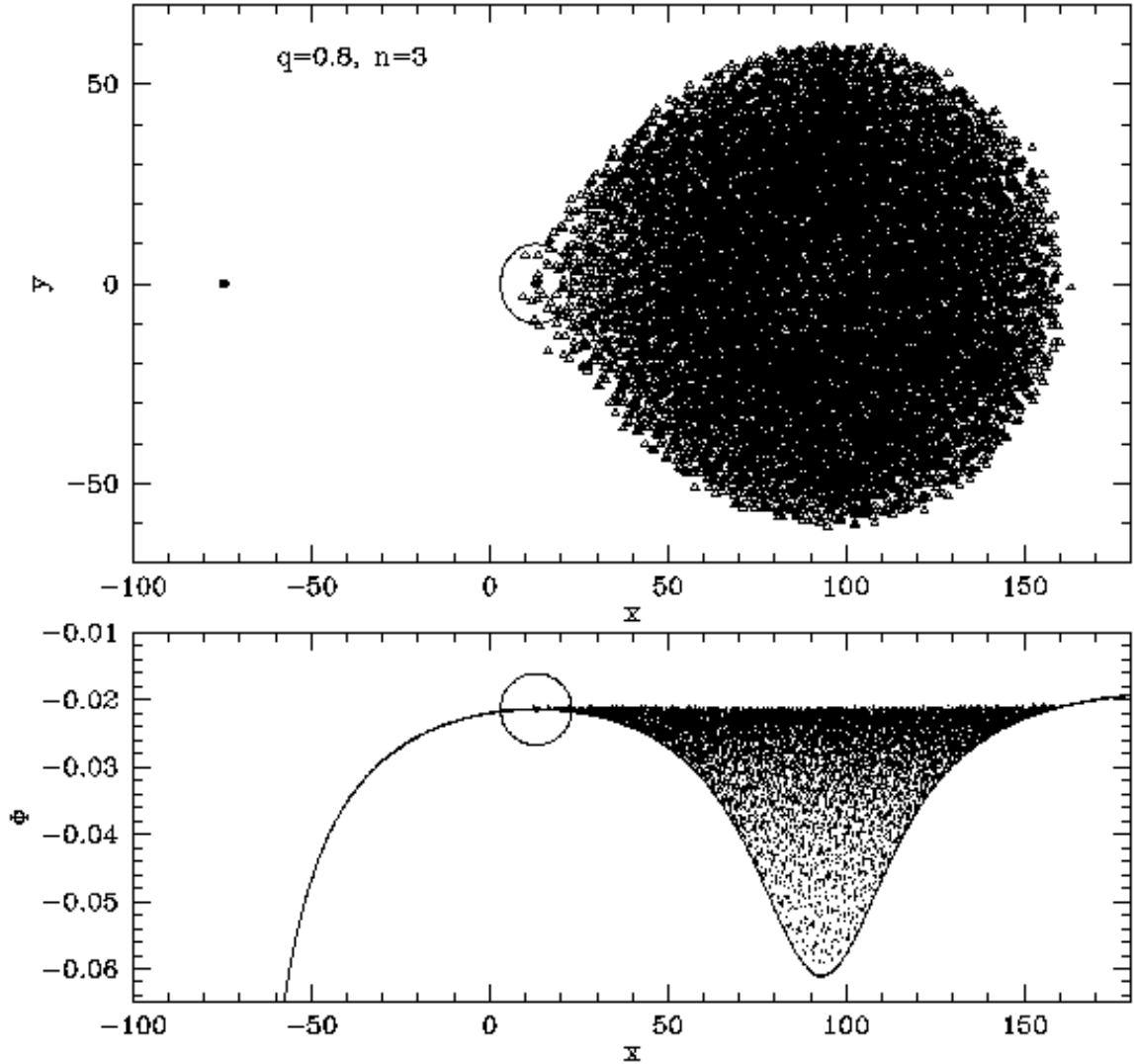,height=160mm,width=160mm} 
\caption{Configuration of 
a polytrope $N$ = 3 and mass ratio $q$ = 0.8
as it fills its critical lobe, {\it i.e} at the onset of particle transfer
towards the primary.
The upper panel is a projection of the SPH particles
onto the orbital $x-y$ plane, $z$ being the rotation axis.
$x$ is the coordinate along the binary axis.
The primary is indicated on the left side of
the plot by a thick dot.
The lower panel displays the projection onto the
($x$, $\Phi$) plane, where $\Phi$ is the effective potential {\it i.e}
the sum of
the gravitational and the centrifugal potentials
(rotating reference frame). 
The solid line indicates the variation of $\Phi(x,y=0,z=0)$ along
the binary axis. In both panels, the inner Lagrangian point is
indicated by a symbol surrounded by an open circle.
The simulation is done with 15000 particles.
}
\label{fig1}
\end{figure*}

\begin{figure*}
\psfig{file=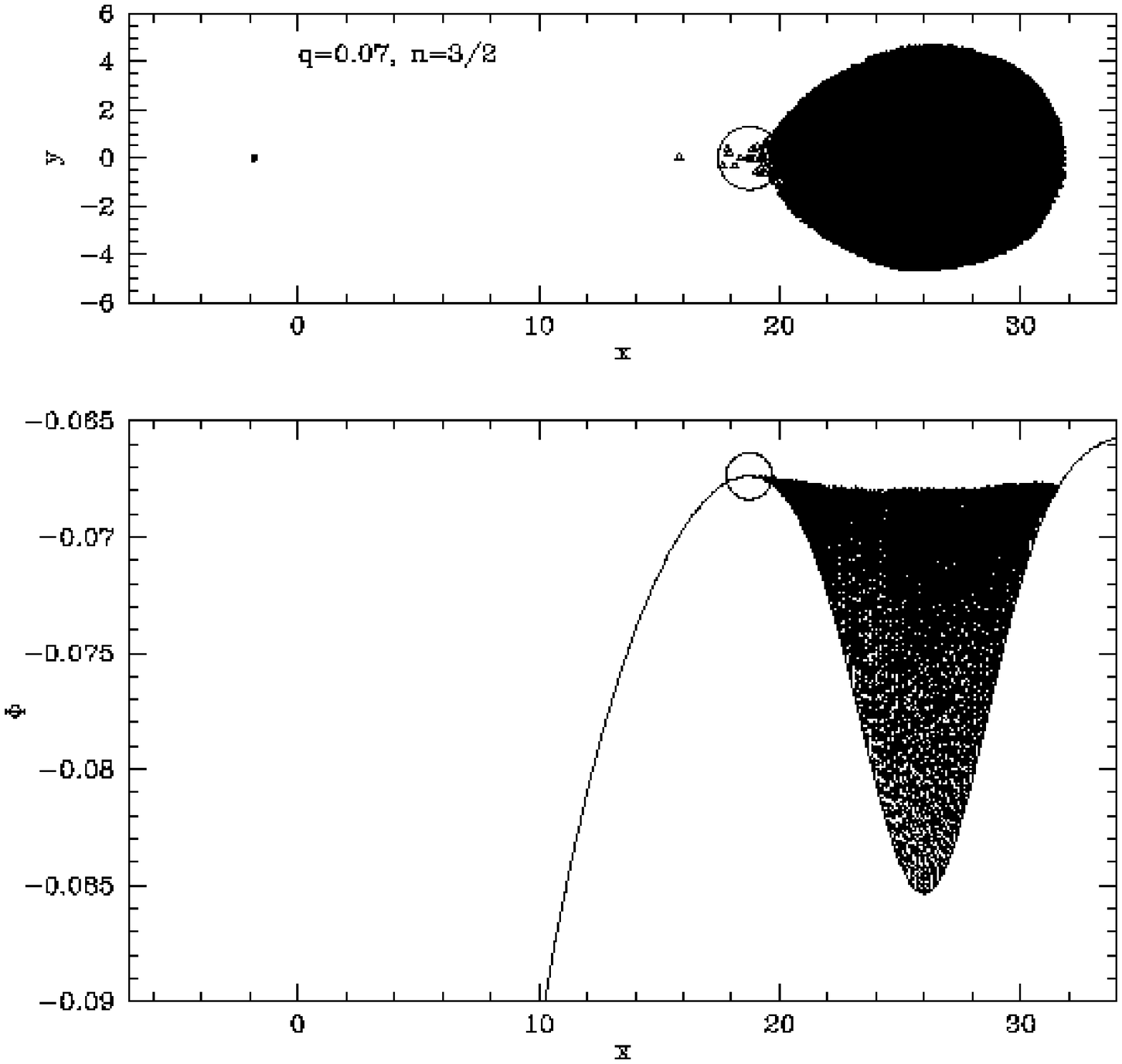,height=160mm,width=160mm} 
\caption{Same as Fig. \ref{fig1} for a $N$ = 3/2 polytrope
and mass ratio $q$ = 0.07. The simulation is done with 57000 particles.
}
\label{fig2}
\end{figure*}

We ran a grid of simulations for various mass ratios $q=M_2/M_1$ between the
secondary and the primary. Typically, CV systems with periods
from $\sim$ 10 h down to the minimum period
cover a range of $q$ between 1 and 0.06. Fig. \ref{fig1} displays the final
configuration of a $N$ = 3 polytrope with mass ratio $q$ = 0.8. 
This illustrates the case of  CV systems with periods $\simgr$ 6 h
(see Baraffe \& Kolb 2000). Fig. \ref{fig2} shows the results
for the case $N$ = 3/2 and $q$ = 0.07, characteristic of secondaries
approaching the period bounce $\pturn$ (Kolb \& Baraffe 1999). 
We note that in the
case $N$ = 3/2 (Fig. \ref{fig2}, lower panel), the surface value of $\phi$ is not
constant. We did not 
find any satisfactory explanation for such behavior. 
This feature has already
been noted in some cases by Rasio and Shapiro (1995) and interpreted
in terms of number density of SPH particles being not exactly
constant around the surface of a star with large tidal deformation. 
Increasing the number of particles from 15000 to 57000 and 
double-checking
that the models have reached an equilibrium configuration do not
solve the problem. 
We do not expect that this affects the accuracy of our
final results, since our deformation calculations are in excellent
agreement
with similar calculations by other authors (see below).

The resulting deformations $D$ 
as a function of $q$ are summarized in Table 1 for $N$ = 3 and
$N$ = 3/2. As expected, tidal and rotational distortion yields
an expansion of the secondary's volume  with respect to the unperturbed
spherical configuration. In terms of effective radius, the expansion
is typically  11\% for  $N$ = 3 and 5\% for  $N$ = 3/2.
The dependence of $D$ on $N$ can be understood
in terms of the compressibility $\chi \, = \, 
\partial \log \rho/\partial p$ = $N/(1 + N)/p$,
which is larger for $N$ = 3 than  for $N$ = 3/2.
The larger the compressibility,  the larger the deformation.

   \begin{table}
      \caption{Deformation $D = R_{\rm f}/R_{\rm i}$ as a function
of mass ratio $q \, = \, M_{\rm donor}/M_{\rm accretor}$ of the secondary
in a semi-detached binary for polytropic indices $N=3$ ($D_3$) and
$N=3/2$ ($D_{3/2}$)}
      \label{tab1}
\begin{tabular}{ccc}
\noalign{\smallskip}
\hline\noalign{\smallskip}
$q$ & $D_3$ &  $D_{3/2}$ \\
\hline\noalign{\smallskip}
0.06 & - & 1.06\\
0.07 & - & 1.06 \\
0.1 &- & 1.05\\
0.2 & 1.12 & 1.05 \\
 0.3 & 1.12 &1.05  \\
 0.4 & 1.12 &1.04 \\
 0.5 & 1.12 &1.04 \\
 0.6 & 1.11 &1.04 \\
 0.7 & 1.11 &1.04 \\
 0.8 & 1.11 &1.04 \\
 0.9 & 1.11 &1.04 \\
 1.0 & 1.10 &1.04 \\
 \hline
  \end{tabular}
   \end{table} 

\begin{figure}
\psfig{file=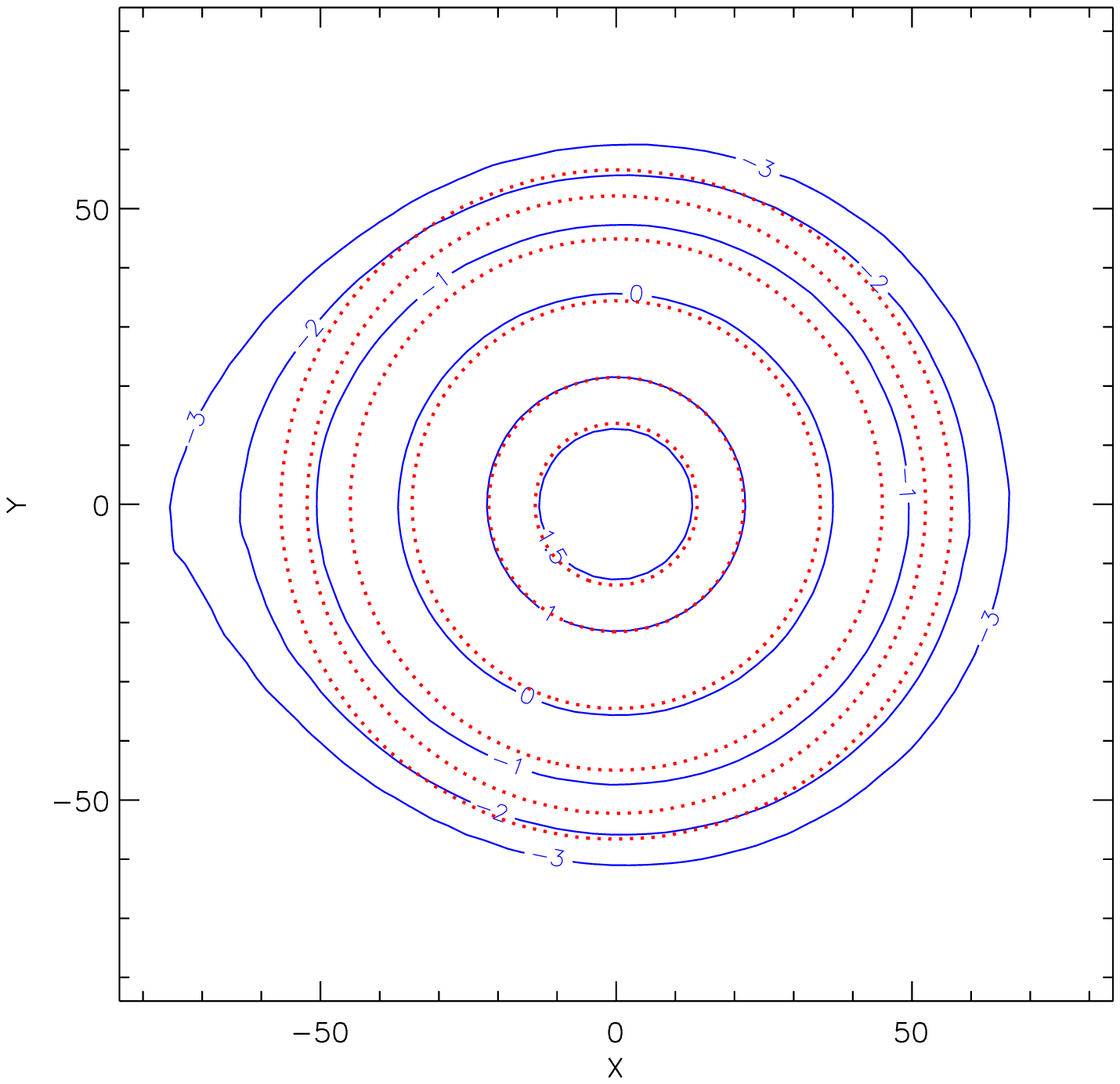,height=88mm,width=110mm}
\psfig{file=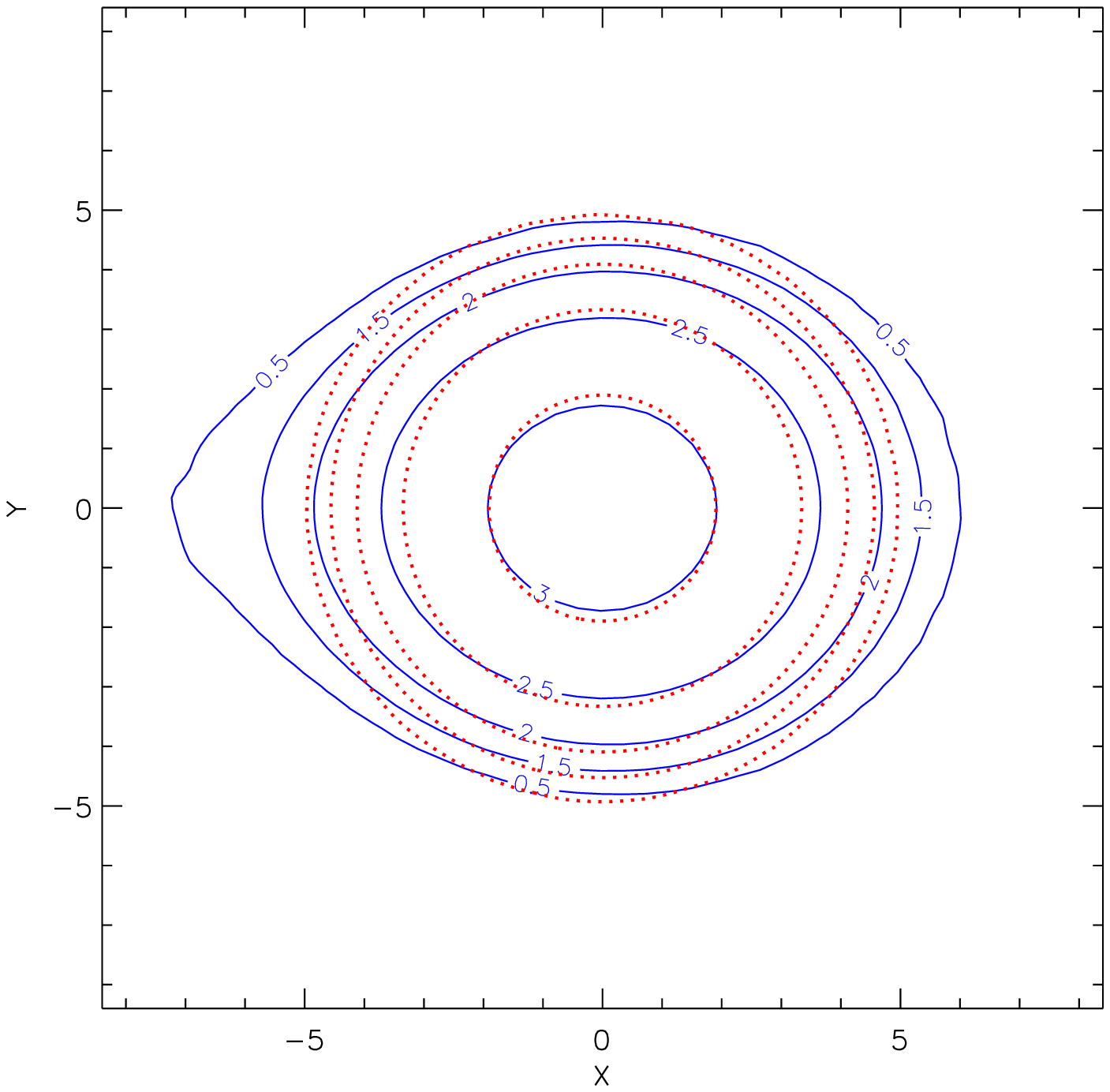,height=88mm,width=110mm} 
\caption{Lines of constant density (in $\log \rho$) of the secondary
in the $x-y$ plane:
dotted lines correspond to the spherical unperturbed secondary
and solid lines to the distorted case. Upper panel: $N$ = 3
and $q$=0.8. Lower panel: $N$ = 3/2 and $q$=0.07. The mass centers
of the spherical and distorted polytropes are located at ($x,y$) = (0,0).
}
\label{fig3}
\end{figure}

In order to visualize the deformation of the secondary
compared to the spherical case, Fig. \ref{fig3} displays lines
of constant density for $N$ = 3 and $N$ = 3/2. 
An inspection of  Fig. \ref{fig3} shows that the largest
departure from spherical symmetry is observed in the outermost
layers of the polytropic star,
whereas the central regions
are only slightly affected.
The results displayed in Table 1 are in excellent agreement with
the work of Uryu \& Eriguchi (1999), based on a different numerical
method. Indeed, for $N$ = 3/2 they found  distortion effects of $\sim$
4\% for 0.1 $\le \, q \, \le $ 1. They however did not analyse the
case $N$ = 3.
A comparison of the numerical ratio $R_{\rm f}/A$,
where $A$ is the orbital separation, and the ratio given within
the Roche model according to Eggleton (1983) shows small differences (less
than 2\%), in agreement with the results of Rezzolla et al. (2001) and
 confirming indeed
that the Roche potential is a good approximation in the present
case (the so-called first assumption, see \S 1).

\section{Application to cataclysmic variable evolutionary models}

In order to analyse  the consequences of
the distortion effects found in the previous section
on  period and mass transfer rate in CV systems, we
follow the secular evolution of the secondary
using the same models and input physics as described in Kolb \& Baraffe
(1999) and Baraffe \& Kolb (2000).
We focus on systems below the 2-3 h period gap and specifically on
the minimum period discrepancy between observations and models. 
Although distortion effects seem to be  more important for systems
above the period gap ($P > 3$ h) 
 (see Tab. \ref{tab1}), their consequences are difficult
to quantify given  the large uncertainties of
evolutionary models describing such
systems, such as the  magnetic
braking law and the resulting mass transfer rate, the evolutionary stage
of the secondary at onset of  mass transfer or the mixing length
parameter.
Below the period gap, such uncertainties are fortunately
considerably reduced
(see Baraffe \& Kolb 2000; Kolb et al. 2001 for details). 

\subsection{Geometrical effects}

In the following, we only consider the effects of distortion
on the geometry of the system. 
The orbital properties, e.g. the orbital
period and separation, and the mass transfer rate will
be indeed affected by the larger effective
radius of the donor, estimated in \S 2, 
compared to the  undistorted case. 
However, for the moment, we ignore the rotational and tidal
effects on the thermal structure of the star, assuming that its
inner structure is unaffected and determined by the unperturbed
stellar structure equations in spherical symmetry.
A rough estimate of the thermal effects on the secondary's
properties resulting from its expansion is derived in the next section
(\S 3.2).

We analyse an evolutionary sequence with an initial donor mass
of 0.21 $\msol$, a primary mass of 0.6 $\msol$, 
and gravitational radiation (GR) as angular
momentum loss mechanism (see Kolb \& Baraffe 1999). 
From the  radius $R_2$ obtained from integration of
the standard stellar structure equations,
and the mass ratio $q$, which varies along the sequence of evolution,
the effective radius is derived  according to
Tab. \ref{tab1}. The mass transfer rate is then calculated as a
function
of the difference between  effective donor radius and Roche radius,
following Ritter (1988).

\begin{figure}
\psfig{file=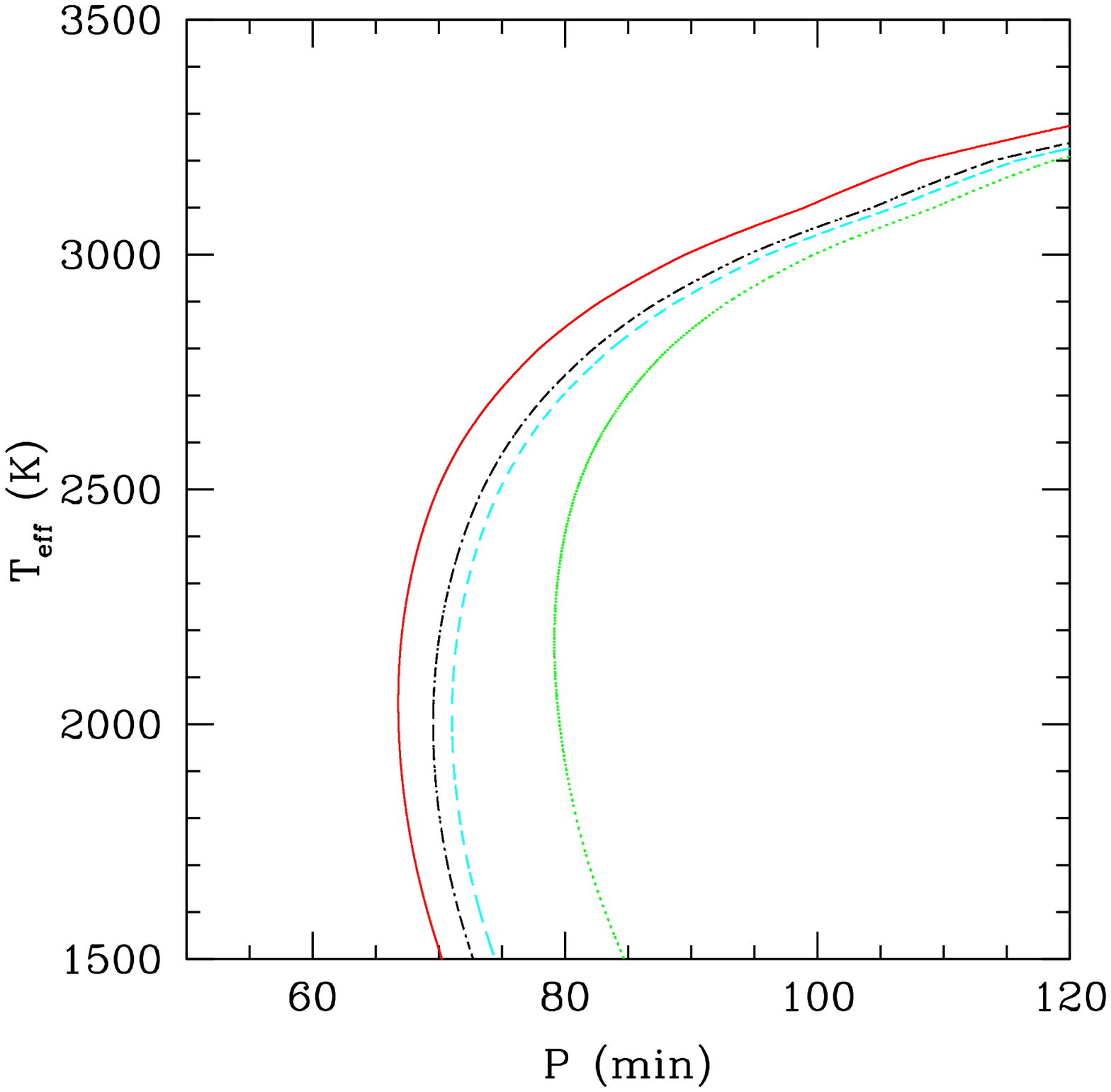,height=110mm,width=88mm} 
\caption{Effective temperature versus orbital period for
evolutionary tracks with undistorted (solid line)
and distorted (dashed line) secondary. 
The dash-dotted line corresponds to the case
with distortion, including thermal relaxation effects as estimated
in \S 3.2.
A test case with distortion
effect included and $\dot J$ = 2.5 $\dot J_{\rm GR}$ is also 
shown (dotted line). 
}
\label{fig4}
\end{figure}

The comparison between sequences without distortion (solid line) and
with distortion (dashed line) is shown in the orbital period -
effective temperature diagram (Fig. \ref{fig4}). Although 
reducing the discrepancy with the observed $P_{\rm min}$,
 distortion effects  provide an increase
of the minimum period $P_{\rm turn}$ of only $\sim$ 6\% (or $\sim$ 4-5 min),
compared to
the undistorted case. 
This is slightly less than what is naively expected from the
period - radius relation $P \propto (R_2^3/M_2)^{1/2}$. An increase of
the radius by $\sim$ 6\%, as expected from distortion effects
near the minimum period (see Tab. \ref{tab1}), should indeed yield
$\sim$ 9\% increase of $P$. The smaller effect found on $P$
stems from the dependence of
angular momentum loss driven by GR on the secondary radius 
$\dot J_{\rm GR}/J\,  \propto \,  P^{-8/3} \,  \propto \, R_2^{-4}$. 
Consequently, the larger radius in the distorted sequence
implies a decrease of $\dot J_{\rm GR}$, and thus   a smaller 
mass transfer rate $-\dot M_2$. 
As shown below $P_{\rm turn}$ depends on the ratio
$\tau=t_{\rm KH}/t_M$ of the secondary's  Kelvin--Helmholtz time
$t_{\rm KH}$ and  the mass transfer time $t_M =
M_2/(-\dot M_2)$. The decrease of $-\dot M_2$ thus yields a decrease of
$\tau$, implying less departure from thermal equilibrium
and thus a smaller $P_{\rm turn}$.  
Because $\dot J_{\rm GR}$ depends explicitly on the mass of the
primary, $P_{\rm turn}$ does also depend on it, but only
 weakly, as  shown by Paczy\'nski \& Sienkiewicz (1983),
who found that $\partial \ln P_{\rm turn}/\partial \ln M_1 \approx
0.09$. In fact our computations show that $P_{\rm turn}$ varies from
71 min for $M_1$ = 0.6 $\msol$ to 74 min for $M_1$ = 1.2 $\msol$, when
distortion effects are included. 

\subsection{Thermal relaxation}

In the numerical computations discussed in the previous section,
we have not taken into account changes in the
thermal reaction of the secondary which must result from its
inflation due to tidal and rotational forces.
A fully consistent treatment of the distortion effects would
imply solving the multi-dimensional stellar structure equations. 
 Rather than doing this we shall in the following derive 
a rough estimate of the thermal relaxation effects and
explore their consequences for the minimum period of CV systems.
We indeed expect that the changed surface area of the more
distended secondary, as a result of the distorsion effects, 
will affect its surface luminosity, and thus its thermal properties.
 

We denote by $R_{2,0}, L_{2,0}, T_{\rm eff, 0}, \dots$ the
quantities of the donor star which result from assuming a pure $1/r$
potential, and by $R_2, L_2, T_{\rm eff}, \dots$ the corresponding
quantities of the spherical equivalent of the critical lobe--filling
star. Obviously we have 
\begin{equation}
\label{tr1}
R_2 = R_{2,0} \, D_{\rm N} \,,
\end{equation}
where $D_{\rm N}$ is the deformation factor for a
polytropic index $N$ determined in \S2.2 (see Table 1).
We thus have for the orbital separation $A$ and the orbital
period $P$ 
\begin{equation}
\label{tr2}
A = A_0 \, D_{\rm N} \,,
\end{equation}
and 
\begin{equation}
\label{tr3}
P =P_0 \, {D_{\rm N}}^{3/2} \,.
\end{equation}
Since mass transfer below the period gap is assumed to be driven by
loss of angular momentum via gravitational radiation alone, the mass 
transfer rate can be written as (see  e.g. Ritter 1996)
\begin{equation}
\label{tr4}
-\dot M_2 = \frac{M_2}{\zeta_{\rm eff} - \zeta_{\rm CL}} \, \left(-2\,
             \frac{\dot J_{\rm GR}}{J}\right) \,.
\end{equation}
Here 
$\zeta_{\rm eff} = d\ln R_2/d \ln M_2$ is the effective mass radius
exponent of the donor star, and $\zeta_{\rm CL}$ the mass radius exponent
of the volume--equivalent critical lobe radius\footnote{$\zeta_{\rm CL}$
is equivalent to the mass radius exponent of the Roche radius
$\zeta_{\rm R}$  in the Roche model (see  e.g. Ritter 1996)}. Because 
\begin{equation} 
\label{tr5}
\frac{\dot J_{\rm GR}}{J} \propto a^{-4}
\end{equation}
we have 
\begin{equation}
\label{tr6}
-\dot M_2 = -\dot M_{2,0} \, {D_{\rm N}}^{-4}\,,
\end{equation}
assuming that $\zeta_{\rm eff}$ is the same. 

Because of the larger radius of the critical lobe--filling star, its
surface is larger,  thereby affecting its luminosity, effective
temperature and Kelvin--Helmholtz time 
\begin{equation}
\label{tr7}
t_{\rm KH} = \frac{G {M_2}^2}{R_2 L_2}  
              = \frac{G {M_2}^2}{4 \pi \sigma {R_2}^3 {T_{\rm eff}}^4}.
\end{equation}
Consequently,  its thermal relaxation, i.e. its gravo--thermal luminosity
$L_{\rm g}$ will also be different.

In order to estimate the change of $t_{\rm KH}$, we need to determine the
change in $T_{\rm eff}$ with radius $R_2$. Since donor stars
below the period gap are fully convective, we can apply the
theory of the Hayashi--line, as described by Kippenhahn and Weigert (1990).
Accordingly  we get 
\begin{eqnarray}
\label{tr8}
\log T_{\rm eff} & = & \frac{3a-1}{5a+2b+5}\, \log R_2 \nonumber \\
                 &   & + \frac{a+3}{5a+2b+5}\, \log M_2 + const. 
\end{eqnarray}
Here $a=(\partial \log \kappa/\partial \log p)_T$ and 
$b=(\partial \log \kappa/\partial \log T)_p$, where $\kappa$ is the
photospheric opacity. Typically, for
very low--mass stars with atmospheric opacities dominated by molecular
absorption, we have
$a \approx 1$ and $b \approx 0$. Therefore 
\begin{equation}
\label{tr9}
\left( \frac{ \partial \log T_{\rm eff}}{\partial \log R_2}\right)_{\rm M}
\approx \frac{1}{5}\, . 
\end{equation}

We can now use (\ref{tr1}) together with (\ref{tr9}) in (\ref{tr7})
and obtain 
\begin{equation}
\label{tr10}
t_{\rm KH} = t_{\rm KH,0} \, {D_{\rm N}}^{-2.8} \,.
\end{equation}
We note that in deriving (\ref{tr10}) we have applied in
(\ref{tr7}) the factor $D_{\rm N}$ only to the $R$--dependence
coming from the luminosity $L_2$. Indeed, the $R_2$ in the
denominator of (\ref{tr7}) comes from the gravitational binding energy
of the star which is related via the Virial theorem to its thermal
energy, i.e. essentially to its central temperature. Since 
the central regions of the star are expected to be hardly affected by 
the distortion of the outer layers (cf. Fig. 3), it is not
appropriate to also propagate the factor $D_{\rm N}$ to the
remaining factor $R_2$. 

Let us now examine the conditions at the period $P_{\rm turn}$. For a
polytope of index $N$ losing mass the effective mass radius exponent can
be written as (e.g. Ritter 1996) 
\begin{equation} 
\label{tr11}
\zeta_{\rm eff} = \zeta_{\rm ad} + \frac{5-N}{3-N}\, 
                  \frac{t_{\rm M_2}}{t_{\rm KH}}\,
                  \frac{L_{\rm g}}{L}\, ,
\end{equation}
where 
\begin{equation}
\label{tr12} 
\zeta_{\rm ad}= \left( \frac{\partial \log R}{\partial \log M}\right)_{\rm K} 
              = \frac{1 - N}{3 - N}
\end{equation}
is the adiabatic mass radius exponent and 
\begin{equation}
\label{tr13}
t_{\rm M_2} = -\frac{M_2}{\dot M_2}
\end{equation}
is the mass loss time scale. Because at $P = P_{\rm turn}$ the
donor star is characterized by $N=3/2$ and $\zeta_{\rm eff}=+1/3$
\footnote{this follows directly from Kepler's 3rd law, the fact that
$(1 +1/q)V_{\rm f}/A^3 \approx {\rm const.}$ for $q \simle 0.8$, and
$\dot P = 0$} we obtain 
from (\ref{tr11}) and (\ref{tr12}) 
\begin{equation}
\label{tr14}
\frac{t_{\rm M_2}}{t_{\rm KH}} \, \frac{L_{\rm g}}{L} =
\frac{2}{7}
\end{equation} 
at $P=P_{\rm turn}$, or 
\begin{equation}
\label{tr15}
\frac{L_{\rm g}}{L}=\frac{2}{7} \,\frac{t_{\rm KH}}{t_{\rm M_2}}
                   \approx \left(\frac{2}{7} \,\frac{t_{\rm KH}}
                    {t_{\rm M_2}}\right)_0 \, 
                    {D_{3/2}}^{-6.8}\,.
\end{equation}
With $D_{3/2} \approx 1.06$ (cf. Table 1) (\ref{tr15}) yields 
\begin{equation}
\label{tr16}
\frac{L_{\rm g}}{L} \approx 0.67 \, 
                            \left(\frac{L_{\rm g}}{L}\right)_0\, .
\end{equation} 
Eq. (\ref{tr16}) means that the 3D--effects, i.e. the reduced mass loss
rate and the increased surface luminosity of the more distended star,
result in a smaller deviation from thermal equilibrium at
$P=P_{\rm turn}$. Hence the star is systematically less inflated by the
effects of thermal disequilibrium, and this, in turn, compensates at
least partially for the systematic increase of the orbital period due
to the factor $D_{3/2} > 1$. 

A quantitative estimate of the decrease of $\pturn$ suggested from
(\ref{tr16}) can be derived by recomputing the evolutionary sequences
including the effect of distortion, as done in  \S 3.1, and
by artificially increasing
the radiating surface of the donor in the Stefan--Boltzmann law by a
factor ${D_{3/2}}^{2.8}$, as suggested from (10). 
Note that this is equivalent to increasing the radiating surface by a
factor ${D_{3/2}}^2$, and  to reducing the surface gravity in the
integration of the stellar atmosphere by the same factor.
 The result of such a numerical experiment is displayed 
in Fig. 4 (dash-dotted line) and 
shows a slight decrease of $P_{\rm turn}$ by $\sim$ 2-3\% compared
to the case with pure geometrical effects (dashed line). These results
fully confirm the expectation derived from (16), namely
that the value of $P_{\rm turn}$ is
reduced by taking into account the effects of the
changed thermal relaxation.  


\section{Discussion and conclusions}

Although reducing the discrepancy between observed
and predicted minimum period, distortion effects seem insufficient
to provide a satisfactory solution of the mismatch between calculated
and observed minimum period. 
A combination of
distortion effects as estimated in \S 2 and  an angular momentum
loss  rate of 2--2.5 $\times \dot J_{\rm GR}$ can reconcile
$P_{\rm turn}$ with 
the observed 80 min value (see dotted line in Fig. \ref{fig4}).
Note that without distortion effects, one would need 4 
$\times \dot J_{\rm GR}$ to reach $\sim 80$ min, as estimated
in Kolb \& Baraffe (1999). The more modest increase of $\dot J$ 
required  according to our calculations is also in better
agreement with Patterson's (1998) estimate
based on space density considerations. 
Additional physical processes can also  result in an inflation
of the secondary, e.g. irradiation from the primary  (Ritter
et al. 2000) or star spots (Spruit \& Ritter 1983).  A rough
estimate of irradiation effects or star spots can be derived by 
following Ritter et al. (2000),  i.e. by reducing the effective
radiating surface of the star by a factor $(1-s_{\rm eff}$). Adopting
in our secular evolution calculation a factor $s_{\rm eff} = 1/2$ and
$\dot J =\dot J_{\rm GR}$ yields a sequence  very similar to the
one obtained with deformation (dashed line in
Fig. \ref{fig4}).  If $s_{\rm eff} = 2/3$  the result
resembles the sequence with deformation and 
$\dot J = 2.5 \times \dot J_{\rm GR}$ (dotted line in Fig. \ref{fig4}).
In order to know  whether such values of $s_{\rm eff}$, yielding
effects comparable to   the distortion effects, are reasonable
requires a sophisticated treatment of  star spots  or
irradiation. 
An investigation of irradiation effects on non-gray stellar 
atmospheres is in progress (Barman 2001).
We stress however that even if distortion, irradiation, star spots or
additional sources of $\dot J$ are possible solutions for
removing the mismatch between observed and predicted minimum period,
the so-called period spike problem still  remains. 
A period spike which is a consequence of  the accumulation
of systems near $P_{\rm turn}$ (where $\dot P = 0$) is indeed
predicted  by all models for which $\dot J$ or $s_{\rm
eff}$  are assumed to be the same for all systems. 
Even if they are not, it is very difficult to ``smear out'' the period
spike in a population of systems with different individual bounce
periods (Barker \& Kolb, in preparation).
Such a period spike is, however, not observed (see
Kolb \& Baraffe 1999).  

Finally, we note that 
Kolb \& Baraffe (1999) obtained negligible effects on the secondary
structure and evolution when applying tidal and rotational corrections
to the 1D stellar structure equations, on the basis of the scheme by
Chan \& Chau
(1979). The effect on the total radius in Kolb \& Baraffe (1999) 
is much smaller ($< 2$\%) than that found from the present
SPH simulations. Since two different numerical methods, on the one hand
the work by Uryu \& Eriguchi (1999) and on the other hand the present
work, predict the same quantitative deformation effects, we are 
confident that the results of our SPH simulation are accurate. 
Although in the SPH
simulations we do not take into account the thermal reaction of the
star to its inflation and deal with polytropes, on the basis of our
simple estimate given in  \S 3.2 we do not expect the thermal effects
to significantly reduce the radius of the deformed star. 
A possible reason for the discrepancy between the calculations by
Chan \& Chau (1979)
and the present results could be the limitation of the former 1D
scheme to describe multi-dimensional effects. 
Figure 3 shows strong effects in the outermost layers
which may be difficult to account for with such a scheme. In any case,
both approaches have their shortcomings, but they both provide the
same conclusion regarding the mismatch of the observed and predicted
minimum period.

To conclude, our SPH simulations suggest that tidal and rotational
distortion effects on the secondary in semi-detached binaries may not
be negligible, and may reach observable levels of $\sim$ 10\% on the
radius for specific cases of polytropic index and mass ratio. Although
this effect yields an increase of the predicted minimum period for CV
systems, it remains too small to explain the observed value of 80 min.
Additional effects such as irradiation, star spots or extra
sources of angular momentum loss still seem to be required, leaving
the problem of the minimum period of CV systems unsettled.

\begin{acknowledgements} We thank W. Benz and  H-C. Thomas for valuable
discussions.  
 I.B thanks the Max-Planck Institut for Astrophysik in Garching
for hospitality during elaboration  of this work.
The calculations were performed using facilities at Centre
d'Etudes Nucl\'eaires de Grenoble.
\end{acknowledgements}

\appendix
\section{Determination of the volume in the SPH calculation}

To estimate the final volume of the secondary at the onset
of mass transfer in our SPH simulations, we proceed as follows. 
We first determine the smallest rectangular box containing
the secondary star. This box with volume $V_{\rm box}$ is then filled 
with $N_{\rm tot}$
points following a Sobol sequence of pseudo-random numbers. Such
a sequence is  self-avoiding, {\it i.e.} the points
are spread out randomly but in a uniform way (see Press et al. 1992),
allowing a more uniform filling of a volume than a standard
random method.
Each point within twice the smoothing length of a particle is
counted, otherwise it is discarded. The number of
points $N_{\rm effective}$ fulfilling such condition determines
the volume $V_{\rm star}$ of the star:

$$ V_{\rm star} =\frac{N_{\rm effective}}{N_{\rm tot}} \times V_{\rm box}. $$

The radius of the star $R_{\rm star}$ is then defined as the radius 
of the sphere
of same volume:

$$ V_{\rm star}=\frac{4}{3} \pi R_{\rm star}^3 $$

This method has been tested on spherical and elliptical structures
of known volumes. According to these tests, it provides
the radius within the size of one particle
$R_{\rm part}$.
Typically, for $N_{part}=15000$, we have

$$ R_{\rm part} \simeq \frac{R_{\rm star}}{N_{\rm part}^{1/3}} \simeq
\frac{R_{\rm star}}{24}. $$

Our method thus determines the radius of a star within a systematic
error of 4\%. Although rough for a precise determination of
a stellar radius, the uncertainty is much smaller on 
the {\it ratio} of the
radius of the same star at two different times of the simulation.
This is the case for the deformation $D$ which
is the main quantity of interest in our analysis. 
If $\epsilon$ is the absolute error on the radius, $R_1$
and $R_2$ the stellar radii at respectively time $t_1$
and $t_2$, one can write:

$$ \frac{R_2 + \epsilon}{R_1 + \epsilon} \simeq
   \frac{R_2}{R_1}\left(1+\epsilon\left(\frac{1}{R_2}-\frac{1}{R_1}
   \right)+o(\epsilon)\right).
$$

For the specific case of distortion calculation, 
 $R_1$ is the
radius of the unperturbed polytrope and $R_2$ the
radius of the polytrope filling its critical lobe.
We find typical values of $R_2/R_1 \leq$ 12\% (see Tables \ref{tab1})
 with $\epsilon \sim$ 4\%. Thus  the first-order term
in our last equation provides a correction of at most 0.5\%. 
Consequently, our  method used for estimating the volume  
is accurate enough for the present study. 
Note that the test
simulations done with 57000 particles give the same distortion
factor (by less than 1\%) than calculations with 15000 particles.

\end{document}